\documentclass[12pt,a4paper,twoside]{article}

\def\TheAuthor{J. Krug}                                   
\def\TheTitle{Population Genetics and Evolution}                        
\def\TheHeading{Population Genetics and Evolution}                                
\def\TheInstitute{Institute for Theoretical Physics}                               
\def\TheAddress{University of Cologne}            
\def\ThePart{F}                                             
\def\TheChapter{2}                                          

\usepackage{ferienschule_18}                                  

\makeindex

\begin{document}

\MakeTitel           
\tableofcontents     

\vfill
\rule{42mm}{0.5pt}\\
{\footnotesize Lecture Notes of the $49^{{\rm th}}$ IFF Spring
School ``Physics of Life''
(Forschungszentrum J{\"{u}}lich, 2018). All rights reserved. }

\newpage


\section{Introduction}

\hspace*{0.5cm} \hfill \textit{Evolution is all about processes that \emph{almost} never happen.} \newline
\hspace*{0.5cm} \hfill Daniel C. Dennett \cite{Dennett} \newline

The mechanism of evolution \iffindex{evolution} by natural selection
\iffindex{natural selection}
was proposed independently by Charles Darwin and Alfred Russell
Wallace in 1858, and constitutes to this day the conceptual framework within which 
the entire mind-boggling diversity of biological phenomena can be organized and, at least in principle, explained. In the famous words of Theodosius Dobzhansky, \textit{``Nothing in biology 
makes sense except in the light of evolution''} \cite{Dobzhansky}. Unlike, for example, the mechanics of Newton or the electrodynamics of Maxwell, the theory of evolution was not originally phrased
in mathematical terms. In fact, a mathematical formulation would not have been possible at the time of Darwin and Wallace, because the nature of heredity was not yet properly understood. 
Although Gregor Mendel published his laws of particulate inheritance already in 1865, they were largely ignored and rediscovered only around 1900. 

At this point the discreteness of the carriers
of genetic information was perceived to be in contradiction to the Darwinian view of evolutionary change being accumulated continuously in small steps over long periods of time. In a striking parallel
to the controversy about the existence of atoms that overshadowed the early days of statistical mechanics \cite{Boltzmann}, 
the proponents of Mendelian genetics were faced with the challenge of explaining how discrete
random changes in the genes of individual organisms could give rise to a continuous and seemingly deterministic evolution of traits on the level of populations and species.  

Not surprisingly, the reconciliation of the two viewpoints required a mathematical formulation of the basic processes through which the genetic composition of a population evolves. This development 
is known as the modern synthesis of evolutionary biology, which
culminated around 1930 in the foundational works of Ronald A. Fisher,
John Burdon Sanderson Haldane and Sewall Wright \cite{Fisher,Haldane,Wright,Wright2}. Wright's summary
of the key insight underlying mathematical population genetics makes it clear why this field may be aptly characterized as a ``statistical mechanics of genes'':

\begin{itemize}

\item[]
\textit{The difficulty seems to be the tendency to overlook the fact that the
evolutionary process is concerned, not with individuals, but with the
species, an intricate network of living matter, physically continuous in space-time,
and with modes of response to external conditions which it appears can
be related to the genetics of individuals only as statistical consequences of
the latter.}~\cite{Wright}

\end{itemize}

Our understanding of the molecular basis of genetic and biological processes has progressed greatly and undergone multiple revolutionary changes since the early days of population 
genetics. Nevertheless the theory  remains fundamental to the interpretation of genomic data in laboratory experiments as well as in natural populations. In this  
lecture some key concepts of mathematical population genetics will be
introduced within the most elementary setting, and a few recent applications 
addressing evolution experiments with bacteria and the evolution of
antibiotic drug resistance will be described. Correspondingly, the
focus will be on asexually reproducing populations; effects of genetic
recombination that play a central role in traditional treatments
of population genetics will not be covered. 
Pointers to the literature for further reading 
are provided throughout the text, and some of the derivations are left as exercises for the reader.

\section{Moran model}
\label{Sec:Moran}

We consider a population consisting of a fixed number $N$ of individuals labeled by an index $i=1,...,N$. Each individual carries a set of hereditary traits which will be collectively referred to as its 
\textit{type}.  Types can change through \textit{mutations} which will however not be explicitly included in the present discussion. 
The most important trait governing the dynamics of type frequencies in
the population is the \textit{fitness} \iffindex{fitness} $f_i$, a real-valued  number which quantifies (in a way to be 
specified shortly) the reproductive success of the individual. In the
Moran model \iffindex{Moran model} individuals reproduce asexually according to the following steps (see Fig.~\ref{Fig:Moran}) \cite{Moran,Nowak,Durrett}:

\begin{itemize}

\item An individual $i$ is chosen for reproduction with a probability proportional to its fitness $f_i$.
\item The chosen individual creates an offspring of the same type. At this point the population size is $N+1$.
\item To maintain the constraint of fixed population size, an individual is chosen randomly (without reference to its fitness) and killed. The killed individual could be the same as the one that was
chosen for reproduction. 

\end{itemize}

\begin{figure}[t]
    \centering
    \includegraphics[width=0.8\hsize]{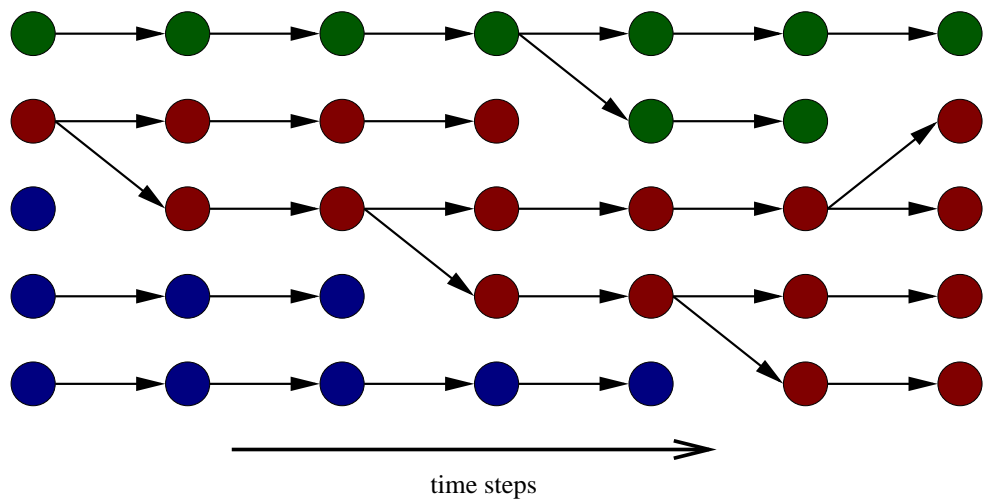}
     \caption{Illustration of the Moran model for a population of size $N=5$. 
     Initially the population consists of three types. After 7 time steps, the blue type has gone extinct and the red type is close to fixation. Note that in the second
     time step the population does not change, because the same individual has been chosen for reproduction and removal.}
     \label{Fig:Moran}
\end{figure}

We say that $N$ of these reproduction events make up one \textit{generation}. The Moran model is one of two commonly used reproduction schemes in population genetics, the other
being the Wright-Fisher model \iffindex{Wright-Fisher model} introduced by two of the pioneers mentioned in the Introduction \cite{Fisher,Wright}. In the Wright-Fisher model the entire population is replaced by its offspring in a single
step. This scheme is advantageous for numerical simulations \cite{PSK2010}, but the Moran model is more easily tractable by analytic means. For large $N$ and under a suitable
rescaling of time the two models are largely equivalent \cite{Blythe,Wakeley}.   

We now specialize to a situation with only two types, denoted by A and B, to which we assign the fitness values $f_A = 1+s$ and $f_B = 1$. The parameter $s$ is called the 
\textit{selection coefficient} \iffindex{selection coefficient} of the A-type relative to the B-type. The state of the population is then 
fully characterized by the number $n_A$ of A-individuals, which we henceforth denote by $n$; the number of $B$-individuals is correspondingly equal to $N-n$. In a single reproduction step
the variable $n$ can change by at most $\pm 1$.  Mathematically speaking the time evolution is a Markov chain on the state space $\{0,1,...,N\}$ governed by the transition probabilities
\begin{equation}
\label{Moran}
T(n+1 \vert n) = \frac{(1+s)n(N-n)}{N^2}, \;\;\;
T(n-1 \vert n) = \frac{n(N-n)}{N^2} 
\end{equation}
and $T(n \vert n) = 1 - T(n+1 \vert n) - T(n-1 \vert n)$, where $T(n'
\vert n)$ denotes the transition probability from $n$ to $n'$. To ensure that the $T(n' \vert n)$ are properly normalized we assume for now that
$s \leq 2$. Markov chains of this kind are also known as birth-death processes \cite{vanKampen}. For later reference we note that by construction
$T(0 \vert 0) = T(N \vert N) = 1$, which implies that $n=0$ and $n=N$ are \textit{absorbing states}. 

\subsection{Selection}
\label{Sec:Selection}

We first ask how the average frequency $x=n/N$ of the A-type changes over time. According to (\ref{Moran}), the average number of A-individuals changes by $\Delta n = sx(1-x)$
in one reproduction step. Counting time in units of generations we thus arrive at 
\begin{equation}
\label{Selection}
\frac{dx}{dt} = \dot{x} = s x(1-x).
\end{equation}
This dynamical system has fixed points at $x=0$ and $x=1$. For $s > 0$ the fixed point at $x=1$ ($x=0)$ is stable (unstable) and the stability is reversed for $s < 0$. Thus under the deterministic
dynamics (\ref{Selection}) the type with higher fitness dominates the population at long times, provided it was at all present initially. This can be seen as a simple mathematical representation
of the Darwinian principle of the survival of the fittest.

It is instructive to rewrite Eq.~(\ref{Selection}) in terms of the mean population fitness $\bar{f} = f_A x + f_B(1-x) = 1+sx$, which yields
\begin{equation}
\label{Replicator1}
\dot{x} = (f_A - \bar{f}) x.
\end{equation}
Thus the frequency of type A grows (shrinks) whenever the mean population fitness is smaller (larger) than $f_A$. Equation (\ref{Replicator1}) can be naturally generalized
to an arbitrary number $K$ of types with fitnesses $f_1,...,f_K$ and
frequencies $x_1,...,x_K$, where it takes the form 
\begin{equation}
\label{Replicator2}
\dot{x_k} = (f_k - \bar{f})x_k, \;\;\;\; \bar{f} = \sum_{k=1}^K f_k x_k.
\end{equation}
This is a simple example of a class of dynamical systems known as
\textit{replicator equations} \iffindex{replicator equation} \cite{Hofbauer}. 

\begin{itemize}

\item[] \textbf{Exercise:} Find the general solution of (\ref{Replicator2}) for arbitrary fitness values $f_k$.
\newline  \textit{Hint:} Look for a transformation that eliminates the nonlinearity
$\bar{f} x_k$.

\end{itemize}

As an immediate consequence of (\ref{Replicator2}), the mean fitness of the population evolves as 
\begin{equation}
\label{Fisher}
\dot{\bar{f}} = \sum_{k=1}^K f_k (f_k - \bar{f})x_k = \sum_{k=1}^K f_k^2 x_k -(\bar{f})^2 = \mathrm{Var}[f] \geq 0,
\end{equation}
where $\mathrm{Var}[f]$ is the population variance of the
fitness. This is a simple version of a statement known as
\textit{Fisher's fundamental theorem} 
\iffindex{Fisher's fundamental theorem} \cite{Fisher}:
The mean population fitness increases under selection, and the rate of fitness increase is proportional to the amount of genetic variability in the population. Within the
framework of the replicator equations  (\ref{Replicator2}) with constant fitness values, selection comes to an end when the fittest type takes over and $\mathrm{Var}[f] \to 0$.
   
\subsection{Drift}

If the two types A,B have the same fitness, $s=0$, then the frequency of the A-type does not change on average. However, the size of the A-population still changes randomly 
according to the transition rates (\ref{Moran}). These fluctuations,
which ultimately arise from the sampling process in a finite
population, are referred to as 
\textit{genetic drift} \iffindex{genetic drift}. 
They are not visible in the deterministic selection equation (\ref{Selection}), which is rigorously valid in the limit $N \to \infty$, because they occur on a different (longer) time scale. 

To study drift we need to refine our analysis and consider the full
distribution $P_n(t)$ of the random variable $n$, which evolves
according to the master equation \cite{vanKampen}
\begin{equation}
\label{Master}
\dot{P}_n(t) = T(n \vert n+1)P_{n+1}(t)+T(n \vert n-1) P_{n-1}(t) - [T(n+1 \vert n) + T(n-1 \vert n)] P_n(t).
\end{equation}
Replacing $n$ by $Nx$ and expanding the transition rates and the distribution function in  powers of $1/N$ one arrives at an evolution equation for the distribution of the A-type frequency
$x$ which takes the form of a Fokker-Planck equation \cite{vanKampen},
\begin{equation}
\label{Diffusion}
\frac{\partial}{\partial t} P(x,t) = - \frac{\partial}{\partial x} s x(1-x) P + \frac{1}{N} \frac{\partial^2}{\partial x^2} x(1-x) P
\end{equation}
up to corrections of order $N^{-2}$. The fact that the diffusion term is of order $N$ implies that \textit{drift phenomena occur on a time scale of $N$ generations.} Provided $s \gg 1/N$ the 
selection term dominates the evolution and one speaks of
\textit{strong selection}\iffindex{strong selection}. Since selection
coefficients are often small, the regime of \textit{weak selection}
\iffindex{weak selection}where
$s \sim 1/N$ is also of importance.

\begin{itemize}

\item[] \textbf{Exercise:} Derive Eq.~(\ref{Diffusion}) from
  Eq.~(\ref{Master}). Can you find stationary solutions of
  Eq.~(\ref{Diffusion})? 
What boundary conditions should you impose at $x=0$ and $x=1$? Are the solutions normalizable?

\end{itemize}

\subsection{Fixation}
\label{Sec:Fixation}

We return to the discrete Markov chain governed by the transition rates (\ref{Moran}). Since the states $n=0$ and $n=N$ are absorbing, for long times the process will reach one of them and subsequently
stay there. When this happens we say that \textit{fixation}
\iffindex{fixation} has occurred at the A-type (if $n=N$) or the B-type (if $n=0$). The probability of fixation at either of the two types depends on the
starting value of $n$. We denote by $\pi_n$ the probability of fixation at the A-type when there were $n$ A-individuals present initially. This quantity can be computed recursively subject to the
obvious boundary conditions 
\begin{equation}
\label{Boundary}
\pi_0 = 0, \;\;\; \pi_N = 1.
\end{equation}
To set up the recursion we consider the evolution of the process starting from $n$. After one time step there are three possibilities:
\begin{itemize}
\item The process has moved to $n+1$ and subsequent fixation occurs with 
probability $\pi_{n+1}$.
\item The process has moved to $n-1$ and subsequent fixation occurs with probability $\pi_{n-1}$.
\item The process stays at $n$ and  subsequent fixation occurs with probability $\pi_{n}$.
\end{itemize}

Summing the three possibilities weighted with their respective transition probabilities we arrive at the relation
\begin{equation}
\label{Recursion}
\pi_n = T(n+1 \vert n) \pi_{n+1} + T(n-1 \vert n) \pi_{n-1} + T(n \vert n) \pi_n
\end{equation}
which defines a second order recursion relation for $\pi_n$.  It is worth noting that the terms on the right hand side of (\ref{Recursion}) are subtly different from those on the right hand side of
the master equation (\ref{Master}). This is because the fixation probability is an eigenvector of the \textit{adjoint} of the time evolution operator of the process, which encodes the dynamics
backwards in time.  The solution of (\ref{Recursion}) which satisfies the boundary conditions (\ref{Boundary}) reads
\begin{equation}
\label{Fixation}
\pi_n = \frac{1 - (1+s)^{-n}}{1-(1+s)^{-N}}.
\end{equation}

\begin{itemize}

\item[]\textbf{Exercise:} Derive Eq.~(\ref{Fixation}) from Eqs.~(\ref{Recursion}) and (\ref{Boundary}).

\end{itemize} 

We next examine some limiting cases of Eq.~(\ref{Fixation}). 

\par\textit{Neutral evolution.} \iffindex{neutral evolution} When $f_A = f_B$ all types have the same fitness and changes in type frequency arise purely by genetic drift. This situation is referred to as
\textit{neutral} evolution \cite{Kimura}. Taking the limit $s \to 0$ in (\ref{Fixation}) we obtain
\begin{equation}
\label{Neutral}
\pi_n = \frac{n}{N},
\end{equation}
a result that can be derived more intuitively from the following argument.  Suppose that initially every individual is of a distinct type but all types have the same fitness.
As time progresses more and more types go extinct until, after a time of the order of the drift time $N$, a single type remains. This implies that one individual in the 
initial population is destined to become the common ancestor of the entire population in the far future. Since all individuals are equivalent under neutral evolution, the probability
for any given individual to acquire this role is $1/N$. For the situation with two types it follows that the probability that the future population will be dominated by the A-type is
equal to the fraction of individuals that are initially of type A, i.e. $n/N$.  

\par\textit{Weak selection.} Taking the joint limit $N\to\infty, n \to \infty, s\to 0$ in (\ref{Fixation}) at fixed values of $x=n/N$ and $\bar{s} = sN$ we arrive at the expression
\begin{equation}
\label{Kimura}
\pi(x) = \frac{1 - \mathrm{e}^{-\bar{s} x}}{1- \mathrm{e}^{-\bar s}}
\end{equation}
which can also be derived from the backwards (adjoint) version of the Fokker-Planck equation (\ref{Diffusion}) \cite{Kimura2}. 

\par\textit{Strong selection.} We specialize to the case of a single initial A-individual which has arisen through a mutation, and take the limit $N \to \infty$ at fixed $s \neq 0$. Then 
(\ref{Fixation}) reduces to
\begin{equation}
\label{SSWM}
\pi_1 = \max\left[\frac{s}{1+s}, 0\right] \approx \max[s,0]
\end{equation}
independent of $N$, where in the last step we have assumed that $s>0$ is small in absolute terms, $0 < s \ll 1$. The biological significance of Eq.~(\ref{SSWM}) 
is that newly arising \textit{deleterious} mutations with $s < 0$ cannot fix in large populations, whereas \textit{beneficial} mutations with $s > 0$ fix with a probability that is proportional
to $s$.  This important  result was first derived by Haldane for the Wright-Fisher model, where $\pi_1 \approx 2s$ for $0 < s \ll 1$ \cite{Haldane}. Quite generally one expects that 
$\pi_1 \sim s$ in this limit with a prefactor that depends on the precise reproduction scheme. Since the selection coefficients of beneficial mutations are rarely larger than a few percent,
this implies that, contrary to the scenario suggested by the deterministic selection equation (\ref{Selection}), a large fraction of beneficial mutations is lost due to genetic drift.  

For later reference it is also of interest to know the time until
fixation, conditioned on it taking place. It turns out that, for the case of strong
selection, the correct result for large $N$ can be obtained from
the deterministic dynamics (\ref{Selection}). Starting the process
with a single A-individual implies that the initial frequency is
$x_\mathrm{initial}=1/N$, and we say that fixation has occurred when
$x = x_\mathrm{final} = 1 - 1/N$. Thus the time to fixation is given
by 
\begin{equation}
\label{tfix}
t_\mathrm{fix} = \int_{x_\mathrm{initial}}^{x_\mathrm{final}}
\frac{dx}{sx(1-x)} = \frac{2 \ln(N-1)}{s},
\end{equation}
which agrees to leading order in $N$ with the full stochastic
calculation. The main contributions to the integral
come from the boundary layers where the frequency $x$ is close to 0 or
1. In this regime the stochastic dynamics is well approximated by a
branching process, which behaves similar to the deterministic
evolution when conditioned on survival \cite{Durrett}. 

\subsection{Parallel evolution}

\iffindex{parallel evolution}
As a simple application of the fundamental result (\ref{SSWM}) we ask for the probability that two populations evolving from the same starting configuration will fix the same mutation. 
We assume that the populations have access to a repertoire of $m$
beneficial mutations with positive selection coefficients $s_1, s_2, ..., s_m$. Then it is plausible (and can be proved
more formally \cite{Gillespie}) that the probability $q_k$ that the first mutation to fix is the one with label $k$ is given by the fixation probability $\pi_1(s_k)$ normalized by the 
sum of all fixation probabilities of beneficial mutations, 
\begin{equation}
\label{qk}
q_k = \frac{\pi_1(s_k)}{\sum_{i=1}^m \pi_1(s_i)} \approx \frac{s_k}{\sum_{i=1}^m s_i}
\end{equation}
under the approximation (\ref{SSWM}).  The probability that the $k$'th mutation is fixed in two independent populations is $q_k^2$. Correspondingly, the probability that any one of the available
mutations fixes in both populations is obtained by summing over $k$, yielding the expression 
\begin{equation}
\label{P2}
P_2 = \sum_{k=1}^m q_k^2 = \sum_{k=1}^m \frac{s_k^2}{\left(\sum_{i=1}^m s_i \right)^2}
\end{equation}
for the probability of parallel evolution \cite{Orr}. 

\begin{figure}[ht]
    \centering
     \includegraphics[width=0.6\hsize]{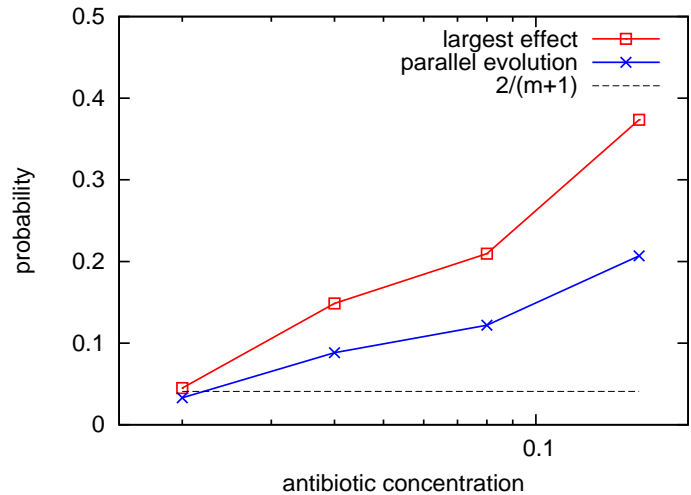}
     \caption{Quantification of parallelism in the evolution of
       antibiotic resistance. Selection coefficients were estimated as
       a function of antibiotic concentration for a panel of 48 mutations in the
       enzyme TEM-1 $\beta$-lactamase that increase resistance against
       cefotaxime. Blue crosses show the probability of
       parallel evolution (\ref{P2}) and red squares the probability $P_\mathrm{max}$
       that the mutation of largest effect size fixes. The horizontal
       dashed line shows the prediction for fitness values distributed
       according to the Gumbel class of extreme value
       theory. Cefotaxime concentration is measured in $\mu$g/ml. Adapted from \cite{Schenk}.}
     \label{Fig:Parallel}
\end{figure}

Similar to entropy measures, $P_2$ quantifies the deviation of the discrete normalized probability distribution $q_k$ from the equidistribution. If all selection coefficients are the same 
$P_2$ takes on its minimal value $1/m$, and any variation in the selection coefficients increases the probability of parallel evolution. In order to actually compute $P_2$ one needs
to invoke empirical data or make specific assumptions about the distribution of selection coefficients.  
Assuming that the tail of the fitness distribution conforms to the Gumbel class of extreme value theory \cite{EVT}, which comprises many common distributions like
the normal or exponential distributions, Orr showed that $P_2$ takes
the universal value $2/(m+1)$ \cite{Orr}. An application to data
obtained in the context of antibiotic resistance evolution
\iffindex{antibiotic resistance} is
shown in Fig.~\ref{Fig:Parallel}.  Here $P_2$ increases systematically with the concentration of antibiotics and considerably exceeds the prediction for the Gumbel class, indicating
that the distribution of selection coefficients is heavy-tailed
\cite{Schenk}. In addition to $P_2$, the figure also shows the
probability $P_\mathrm{max} = \max_k q_k$ that the mutation of largest
effect fixes first. 

\begin{itemize}

\item[] \textbf{Exercise:} Prove that $P_\mathrm{max} \geq P_2$.

\end{itemize}

\section{Regimes of evolutionary dynamics}
\label{Sec:Regimes}

We make use of the results derived in the preceding section to
describe some simple but biologically relevant regimes of evolutionary
dynamics. 

\subsection{Molecular clock}

\iffindex{molecular clock}
Suppose that \textit{neutral} mutations arise in the population at rate $U_n$
per individual and generation. The average number of mutations arising
in the whole population per generation is then $U_n N$, a fraction $\pi_1 = 1/N$ of
which go to fixation according to Eq.~(\ref{Neutral}). Thus the
rate at which nucleotide changes occur in the genetic sequence,
referred to as the (neutral) \textit{substitution rate}
\iffindex{substitution rate} 
\begin{equation}
\label{Substitution}
\nu_n = \pi_1 U_n N = U_n
\end{equation} 
is independent of population size. This simple observation underlies
the molecular clock hypothesis, which posits that molecular evolution 
at the level of amino acid substitutions in proteins occurs at approximately
constant rate across widely different species \cite{Kumar}.  
 
\subsection{Clonal interference}
\label{Sec:CI}

When a population is not optimally adapted to its environment, a
certain fraction of mutations are beneficial. Suppose that such 
mutations occur at rate $U_b$ per individual and generation, and that
the associated typical selection coefficients satisfy 
$N^{-1} \ll s \ll 1$. Then the
rate of substitution is $\nu_b = \pi_1 U_b N = s U_b N$ which
increases linearly in $N$. Stated differently, the time between substitutions
\begin{equation}
\label{tsub}
t_\mathrm{sub} = \frac{1}{\nu_b} = \frac{1}{N s U_b}
\end{equation}
decreases with increasing population size. On the other hand, the time 
(\ref{tfix}) required for a single fixation event increases
logarithmically in $N$. As a consequence, we need to distinguish two
regimes of adaptive evolution that occur at small vs. large population sizes. 
 
\begin{figure}[ht]
    \centering
     \includegraphics[width=0.75\hsize]{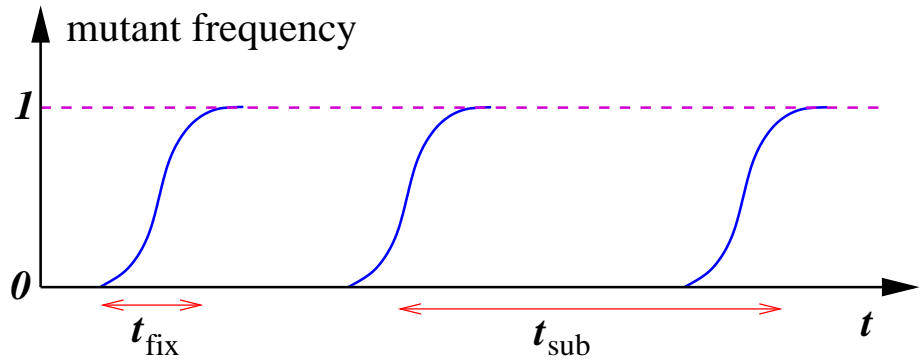}
     \caption{Illustration of the periodic selection
       regime. Beneficial mutations that overcome genetic drift emerge
     at the substitution rate $\nu_b = 1/t_\mathrm{sub}$ and fix in
     independent selective sweeps.}
     \label{Fig:Periodic}
\end{figure}

\par In the \textit{periodic selection regime} \iffindex{periodic selection}($t_\mathrm{sub} \gg
t_\mathrm{fix}$) each mutation has time to fix before the next one
appears, and fixation events are independent
(Fig.~\ref{Fig:Periodic}).  As $N$ increases, it becomes increasingly likely that fixation events overlap, in the sense that a second beneficial mutation arises 
while the first is still on the way to fixation. Because the second mutation increases the mean fitness of the population, it reduces the selective advantage 
of the first mutation [compare to (\ref{Replicator2})]. Through this mechanism multiple beneficial clones that are present simultaneously in the population 
compete for fixation, a phenomenon known as \textit{clonal
  interference} 
\iffindex{clonal interference} \cite{PSK2010,Gerrish,ParkKrug2007}. 
Clonal interference is the dominant mode of evolution
when $t_\mathrm{sub} \ll t_\mathrm{fix}$ or 
\begin{equation}
\label{CIcondition}
\frac{t_\mathrm{fix}}{t_\mathrm{sub}} = 2 N U_b \ln N \gg 1.
\end{equation}
This condition is often satisfied for populations of bacteria or viruses, which are characterized by large population sizes and large mutation rates. As an example, we consider the famous
long-term evolution experiment with \textit{Escherichia coli} that was initiated by Richard Lenski in 1988. The experiment started with 12 genetically identical populations in a minimal
growth medium, which allows the bacteria to grow but leaves room for adaptation. Each day the populations are transferred to fresh growth medium following a fixed protocol, and population
samples are stored in a freezer every 500 generations. In this experiment the beneficial mutation rate has been estimated to be $U_b \approx 1.7 \times 10^{-6}$, and the population
size (averaged over serial transfers) is $N = 3.3 \times 10^7$ \cite{Wiser}. With these numbers the quantity on the right hand side of the criterion (\ref{CIcondition}) is about 2000,
and indeed a recent genomic analysis of the first 60000 generations in the Lenski experiment reveals massive clonal interference \cite{Good}.    

\subsection{Speed of evolution}
\label{Sec:Speed}

Clonal interference implies that beneficial mutations can be outcompeted by others even when they have reached a frequency that is large enough to make genetic drift irrelevant. 
Thus beneficial mutations are lost that would otherwise have fixed, which reduces the speed at which the fitness of the population increases. If the beneficial mutations cover a range of 
selection coefficients then those of small effect are more likely to survive competition, and the distribution of fixed mutations is shifted towards larger effect sizes (Fig.~\ref{Fig:CI}).   

\begin{figure}[ht]
    \centering
    \includegraphics[width=0.45\hsize]{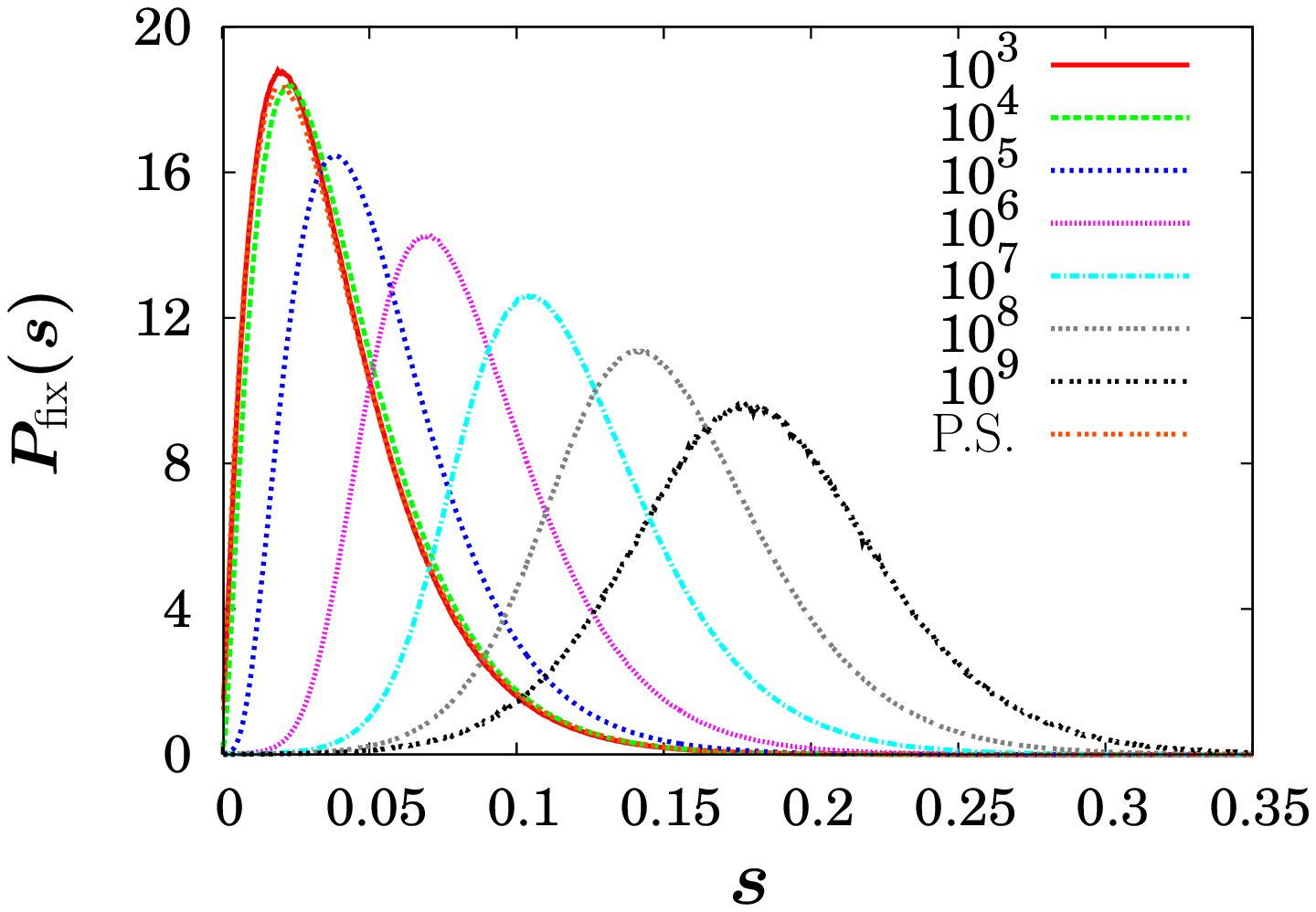} \includegraphics[width=0.45\hsize]{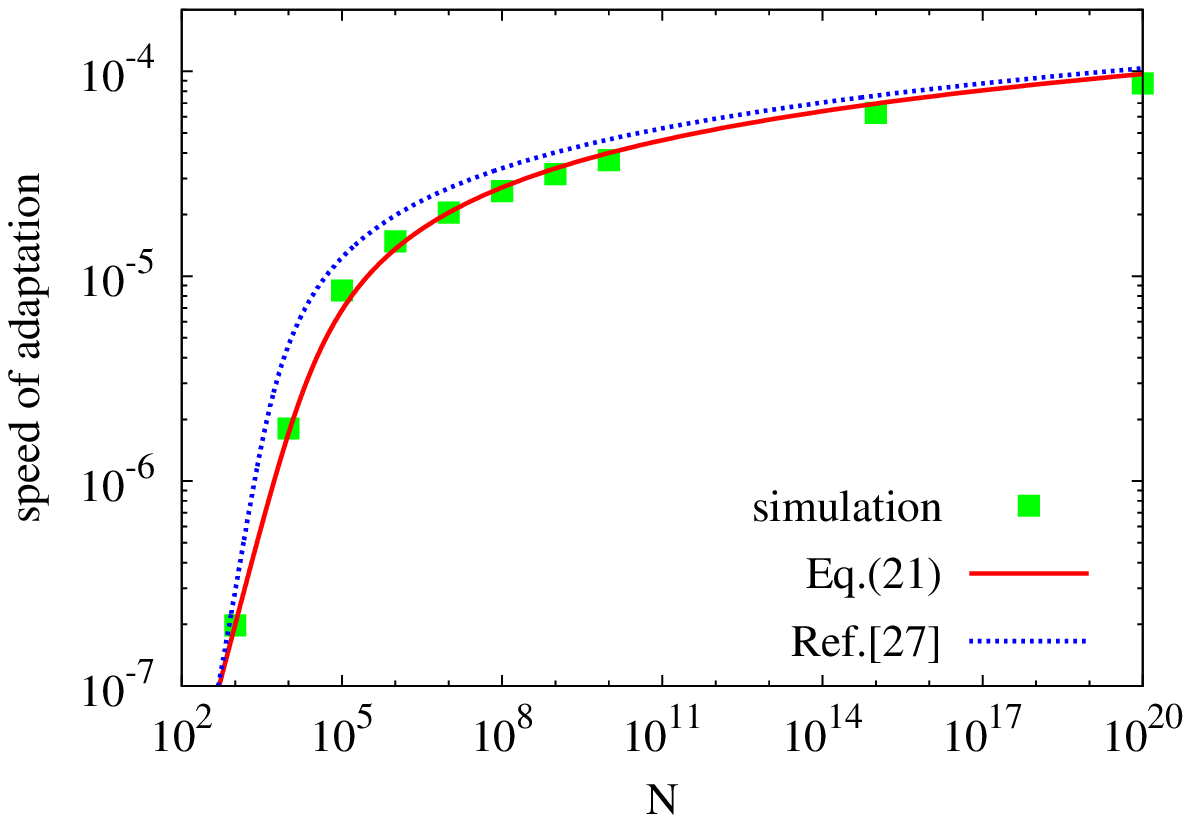}
     \caption{Signatures of clonal interference in simulations of the Wright-Fisher model with beneficial mutations occurring at rate $U_b = 10^{-6}$. 
     Left: Distribution of selection coefficients of fixed mutations
     for different population sizes. In these simulations selection
     coefficients 
     were drawn randomly from an exponential distribution with mean
     $s_b = 0.02$ \cite{ParkKrug2007}. In the periodic selection
     regime (P.S.)
     mutations that survive genetic drift fix independently. Since the
     fixation probability is proportional to $s$, the distribution of
     fixed mutations is $\sim s \mathrm{e}^{-s/s_b}$ 
For $N \geq 10^5$ clonal interference becomes relevant and the
distribution shifts to larger values. Courtesy of Su-Chan Park.
     Right: Speed of adaptation as a function of population size assuming a single beneficial selection coefficient $s = 0.01$. The full red line shows Eq.~(\ref{Speed}), the blue 
     dotted line a related expression derived in \cite{Rouzine}, and the red squares are simulation results. Adapted from \cite{ParkKrug2013}.    
     }
     \label{Fig:CI}
\end{figure}

To quantify the effect of clonal interference on the speed of fitness increase, we assume that all beneficial mutations have the same selection coefficient $s$, irrespective of when
they occur. Then each fixed mutation increases fitness by $s$, and the rate of fitness increase in the periodic selection regime 
\begin{equation}
\label{Vper}
V = V_\mathrm{PS} = s \nu_b = s^2 U_b N
\end{equation}
is proportional to the supply rate $U_b N$ of beneficial mutations. Computing the speed of evolution in the clonal interference regime requires analyzing the complex stochastic dynamics 
of interacting clones, and no simple closed-form solution exists \cite{PSK2010}. An approximate implicit formula for the speed $V(N)$ that covers both regimes is given by \cite{Rouzine,ParkKrug2013}
\begin{equation}
\label{Speed}
\ln N \approx \frac{V}{2 s^2} \left[\ln^2 \left(\frac{V}{e U_b s}\right) + 1 
\right ] + \ln \left(\frac{V}{2s^2 U_b}\right),
\end{equation}
see Fig.~\ref{Fig:CI} for a comparison to simulation data. For large $N$ the speed increases logarithmically rather than linearly with $N$, a behavior that persists up to hyperastronomically
large population sizes \cite{PSK2010}.

\begin{itemize}

\item[] \textbf{Exercise:} Find the solution $V(N)$ of the implicit equation (\ref{Speed}) for small and large $N$. Note that the result for small $N$ differs slightly from (\ref{Vper}), because 
(\ref{Speed}) was derived assuming Wright-Fisher dynamics. 

\end{itemize}

Experiments in which the population size and the beneficial mutation rate are varied systematically are in qualitative agreement with the predictions outlined above
\cite{deVisser1999,deVisser2006}. However, the assumption that beneficial mutations of constant effect size emerge at a constant rate, and that therefore the fitness of the population
increases linearly in time, is not consistent with long-term experiments. Instead, the fitness increase slows down over time, an effect that is attributed to a decrease of the typical
selection coefficient \cite{Wiser}. Since this implies that the mutational effect size depends on the fitness of the population in which it occurs, it constitutes a simple case
of \textit{epistasis} (see Sec.~\ref{Sec:Epistasis} for further discussion).  

\subsection{Muller's ratchet and the evolutionary benefits of sex}
\label{Sec:Muller}

\iffindex{Muller's ratchet} \iffindex{sex}
The large majority of random mutations is deleterious or lethal \cite{Eyre-Walker}. In the preceding subsections we have nevertheless ignored deleterious mutations, because we have
seen in Sec.~\ref{Sec:Fixation} that they cannot fix in large populations. However, even if deleterious mutations are constantly removed from the population by selection, their presence
reduces the fitness of the population, an effect that is called
\textit{mutational load}.\iffindex{mutational load} To quantify it, we use the deterministic framework developed in Sec.~\ref{Sec:Selection}. Suppose deleterious 
mutations with selection coefficient $s = - \tilde s < 0$ arise at constant rate $U_d$, and that an individual with $k$ deleterious mutations has fitness $f_k = - \tilde s k$. There are no 
beneficial or neutral mutations. Then the frequency $x_k$ of individuals with $k$ mutations making up the $k$'th \textit{fitness class} evolves according to \cite{Haigh}
\begin{equation}
\label{Muller}
\dot{x}_k = -\tilde{s}(k-\bar{k}) x_k + U_d \, x_{k-1}-U_d \, x_k \;\;\; \textrm{with} \;\;\; \bar{k} = \sum_k k x_k.
\end{equation}
This set of equations has a stationary solution $\bar{x}_k$ that takes the form of a Poisson distribution
\begin{equation}
\label{Poisson}
\bar{x}_k = \frac{\Lambda^k \mathrm{e}^{-\Lambda}}{k!} \;\;\; \textrm{with} \;\;\; \Lambda = \frac{U_d}{\tilde s}.
\end{equation}
Remarkably, the mutational load $s \bar{k} = -U_d$ is independent of $\tilde s$. 

\begin{itemize}

\item[] \textbf{Exercise:} Verify the stationary solution (\ref{Poisson}) of (\ref{Muller}), and show that there is an infinite number of stationary solutions related to (\ref{Poisson}) by 
an integer shift of $k$. 
   
\end{itemize}   
 
At this point we recall that real populations are finite, and therefore the deterministic description (\ref{Muller})  has to be complemented by the effects of genetic drift.
If the total population size is $N$, the number of individuals in the fittest (mutation-free) class is, on average, 
\begin{equation}
\label{N0}
\bar{n}_0 = \bar{x}_0 N = N \mathrm{e}^{-U_d/\tilde s}.
\end{equation}
In fact the number of mutation-free individuals is a random quantity that fluctuations between $n_0 = N$ and $n_0 = 0$. Because beneficial mutations do not occur, the 
state $n_0 = 0$ is absorbing: Once the mutation-free class has become extinct, it cannot be reconstituted. The time until the (inevitable) extinction of the mutation-free
class occurs depends crucially on its mean size (\ref{N0}). If $\bar{n}_0 \gg 1$, we know from the results of Sec.~\ref{Sec:Fixation} that
the fixation of the deleterious mutants is exponentially unlikely, and hence the time until extinction is exponentially large in $N$. Once it occurs, the frequency distribution
has ample time to relax back to a shifted Poisson distribution for which the minimal number of mutations is now $k=1$. This process repeats itself periodically and is referred
to as one \textit{click of the ratchet}. In contrast, when $\bar{n}_0 \sim 1$ extinction events occur rapidly and fitness declines steadily without discernible discrete clicks.
The computation of the speed of fitness decline as a function of $N$, $U_d$ and $\tilde{s}$ is a difficult problem that is the subject of current research \cite{Rouzine}. 
In the regime $\bar{n}_0 \gg 1$ the dynamics is dominated by rare events, and methods from physics based on the WKB approximation of quantum mechanics have proven
to be useful \cite{Metzger}.

The ratchet-like fitness decline in asexual populations subject to a constant rate of deleterious mutations was first proposed by Hermann Muller as a possible explanation for the 
evolutionary advantage of sexual reproduction \cite{Muller1964}. In sexual populations mutation-free offspring can be created by recombining the genomes of parents which have
deleterious mutations at different positions of the genome, and therefore the ratchet can be halted. It is worth pointing out that also the phenomenon of clonal interference described above 
in Sects.~\ref{Sec:CI} and \ref{Sec:Speed} was first discussed in the context of comparing sexual and asexual reproduction \cite{Fisher,Muller1932}. In sexual populations clonal
interference is alleviated, because beneficial mutations that have originated in different population clones can recombine into a single genome, and therefore sexual reproduction
confers an advantage in terms of the speed of adaptation. Theoretical analysis shows that even a small rate of recombination substantially increases the speed \cite{ParkKrug2013}.  
  
\subsection{Spatial structure}

All models introduced so far assume that the population is {\em well-mixed}, in the sense that the competition induced by the constraint of fixed population size acts globally. This assumption is 
valid when bacteria are grown in shaken liquid culture, but does not apply to the growth of colonies on plates, nor does it generally apply to natural populations. Incorporating an explicit
spatial structure where competition between individuals is implemented locally leads to important changes in the scenarios described above. For example, because a clone of beneficial
mutants can grow only at the boundary of the region that it inhabits, the expression (\ref{tfix}) for the time to fixation is modified. Assuming that the spreading of the clone occurs at a speed
proportional to the selection coefficient and that the population density is constant, one obtains \cite{Martens}
\begin{equation}
\label{tfixspatial} 
t_\mathrm{fix} \sim \frac{l}{s} \sim \frac{N^{1/d}}{s},
\end{equation}
where $l$ denotes the linear extension of the system and $d$ the spatial dimension. Thus fixation is considerably slower than in the well-mixed case, and correspondingly
clonal interference is enhanced. Moreover, instead of the logarithmic dependence on $N$ discussed in Sec.~\ref{Sec:Speed} the speed of adaptation attains a finite speed limit 
for large systems which scales as $V_\infty \sim s^2 U_b^{1/(d+1)}$ \cite{Martens}. Also the dynamics of Muller's ratchet explained in Sec.~\ref{Sec:Muller} is altered significantly, in that the 
population fitness declines at a constant rate even in very large populations, provided the rate of deleterious mutations $U_d$ is large enough \cite{Otwinowski}.
 
\section{Epistasis and fitness landscapes}
\label{Sec:Epistasis}

\iffindex{epistasis}
In the discussion of the dynamical regimes in Sect.~\ref{Sec:Regimes} it was assumed that a mutation can be associated with a selection coefficient that quantifies its effect 
on fitness independent of the type or \textit{genetic background} of the individual in which it occurs. In most cases this is an oversimplification, and \textit{epistatic interactions} between the effects of 
different mutations have to be taken into account
\cite{Phillips,Poelwijk}. In this final section we sketch the
mathematical description of epistasis among mutations occurring at
multiple genetic loci. This will lead to the important concept of a
\textit{fitness landscape}, which defines the arena in which the evolutionary
processes described in the preceding sections ultimately take place \cite{Wright2,deVisserKrug}.  

\subsection{Pairwise interactions}

To fix ideas,  consider a \textit{wild type} with fitness $f_0$ and two different mutant types with fitness $f_1$ and $f_2$, respectively. The selection coefficients of the two mutations are 
$s_1 = f_1 - f_0$ and $s_2 = f_2 - f_0$. If these selection
coefficients were independent of the genetic background we would
expect that the fitness $f_{12}$ of the double mutant is given by
\begin{equation}
\label{Nonepistatic}
f_{12}^{(0)} = s_1 + f_2 = s_2 + f_1 = f_1 + f_2 - f_0.
\end{equation}
Any deviation from this linear relation implies that the two mutations interact epistatically, and the strength and sign of the interactions are quantified by the (pairwise) epistasis coefficient
\begin{equation}
\label{Epsilon}
\epsilon_2 = f_{12} -f_{12}^{(0)} = f_{12} + f_0 - f_1 - f_2.
\end{equation}  
An important qualitative distinction can be made according to whether
the sign of the selection coefficient associated with a given
mutation depends on the genetic background or not; 
in the first case one speaks of \textit{sign epistasis}, in the second
of \textit{magnitude epistasis} \cite{Weinreich2005}. In the example
at hand, the selection coefficient of the second mutation in the
background of the first is $s_2^{(1)} = f_{12} - f_1 = s_2 +
\epsilon_2$, 
and hence its sign is affected by the first mutation if $s_2 s_2^{(1)} < 0$ or
$\epsilon_2 s_2 < -s_2^2$.   
 
\begin{figure}[ht]
    \centering
     \includegraphics[width=0.9\hsize]{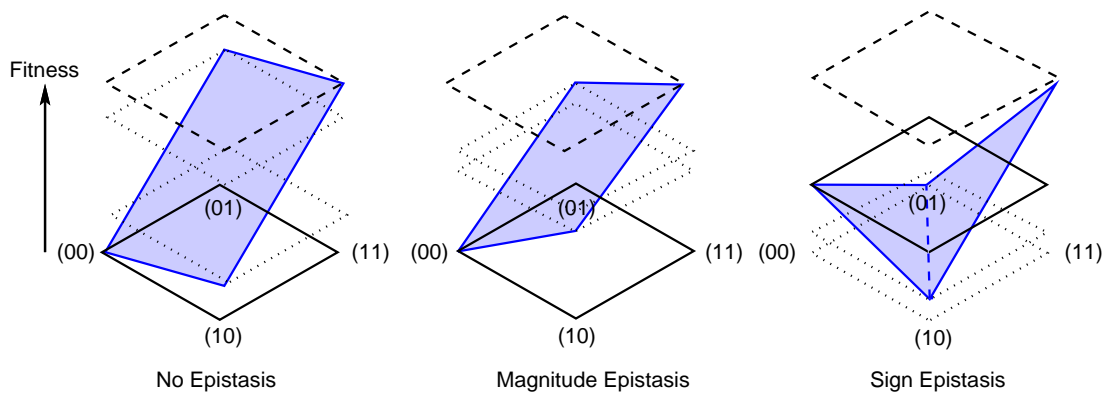}
     \caption{Different types of epistasis between mutations occurring at
       two different genetic loci. The panel on the far right shows a
       case of reciprocal sign epistasis, where the sign of the
       selection coefficient of the first mutation depends on the
       presence of the second and vice versa. Note that the
       corresponding fitness landscape displays two local maxima.}
     \label{Fig:Epistasis}
\end{figure}

Magnitude and sign epistasis for a pair of mutations is illustrated in
Fig.~\ref{Fig:Epistasis}. In this figure the four types are encoded by
a pair $\sigma = (\sigma_1, \sigma_2)$ of binary variables $\sigma_i \in
\{0,1\}$, where $\sigma_i = 0/1$ implies the absence/presence of
the $i$'th mutation. The fitness values of the four types can then be succinctly written in the form
\begin{equation}
\label{2locus}
f(\sigma_1, \sigma_2) = f_0 + s_1 \sigma_1 + s_2 \sigma_2 + \epsilon_2
\sigma_1 \sigma_2.
\end{equation}
Equation (\ref{2locus}) is the simplest example of a fitness
  landscape \iffindex{fitness landscape} which assigns fitness values to a collection of genotypes
\cite{deVisserKrug}. In the present case the genotypes consist
of two genetic loci $i=1,2$, each of which carries two
possible \textit{alleles}  \iffindex{allele}$\sigma_i = 0$ or 1. Because the four
genotypes are located at the corners of a square, the landscape is
easily visualized by arranging the genotypes in a plane and plotting
fitness as the third dimension. This property is lost when we extend
the description to more than two loci. 

\subsection{Multiple loci}

In the general case of $L$ mutational loci the genotype is described
by a sequence of length $L$, $\sigma = (\sigma_1,
\sigma_2,...,\sigma_L)$. In practice there could be more than two
alleles at each site. For example, when describing mutations on the
level of DNA sequences each site carries one of 4 nucleotide bases, and in
proteins the set of alleles are the 20 amino acids. Here we restrict
ourselves to the binary case and let $\sigma_i \in \{0,1\}$ represent
the absence or presence of a specific mutation as before. Then a
generic fitness landscape $f(\sigma)$ can be written in the form
of a discrete `Taylor' expansion \cite{Poelwijk,Szendro}
\begin{equation}
\label{Llocus}
f(\sigma) = f_0 + \sum_{i=1}^L s_i \sigma_i + \sum_{k=2}^L
\sum_{\{i_1,i_2,..,i_k\}} \epsilon_k^{\{i_1,i_2,...,i_k\}} \sigma_{i_1}
\sigma_{i_2}\cdot \cdot \cdot \sigma_{i_k},
\end{equation}   
where the first sum on the right hand side contains the non-epistatic (linear) contributions. 
The second sum runs over
all subsets $\{i_1,i_2,...,i_k\}$ of $k \geq 2 $ out of $L$ loci. The
coefficients $\epsilon_k^{\{i_1,i_2,...,i_k\}}$ generalize the
pairwise epistasis coefficient $\epsilon_2$ in (\ref{2locus}). As
there are $L \choose k$ coefficients of order $L$, the total number of
coefficients in the expansion (\ref{Llocus}) is equal to $2^L$. Thus
the mapping of the $2^L$ fitness values to the expansion coefficients is one-to-one.

\begin{figure}[ht]
    \centering
     \includegraphics[width=0.4\hsize]{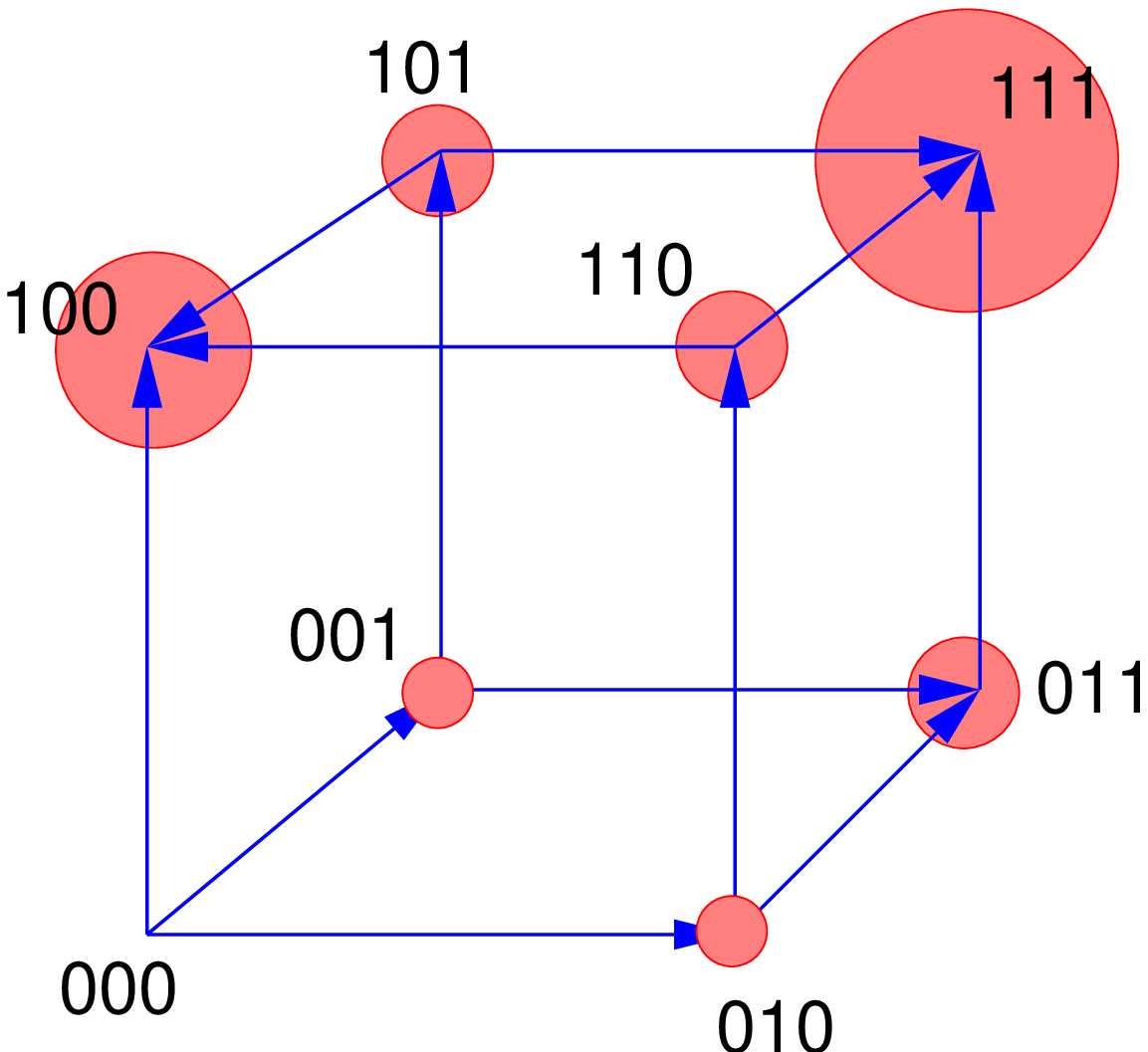} \hspace*{0.5cm} \includegraphics[width=0.5\hsize]{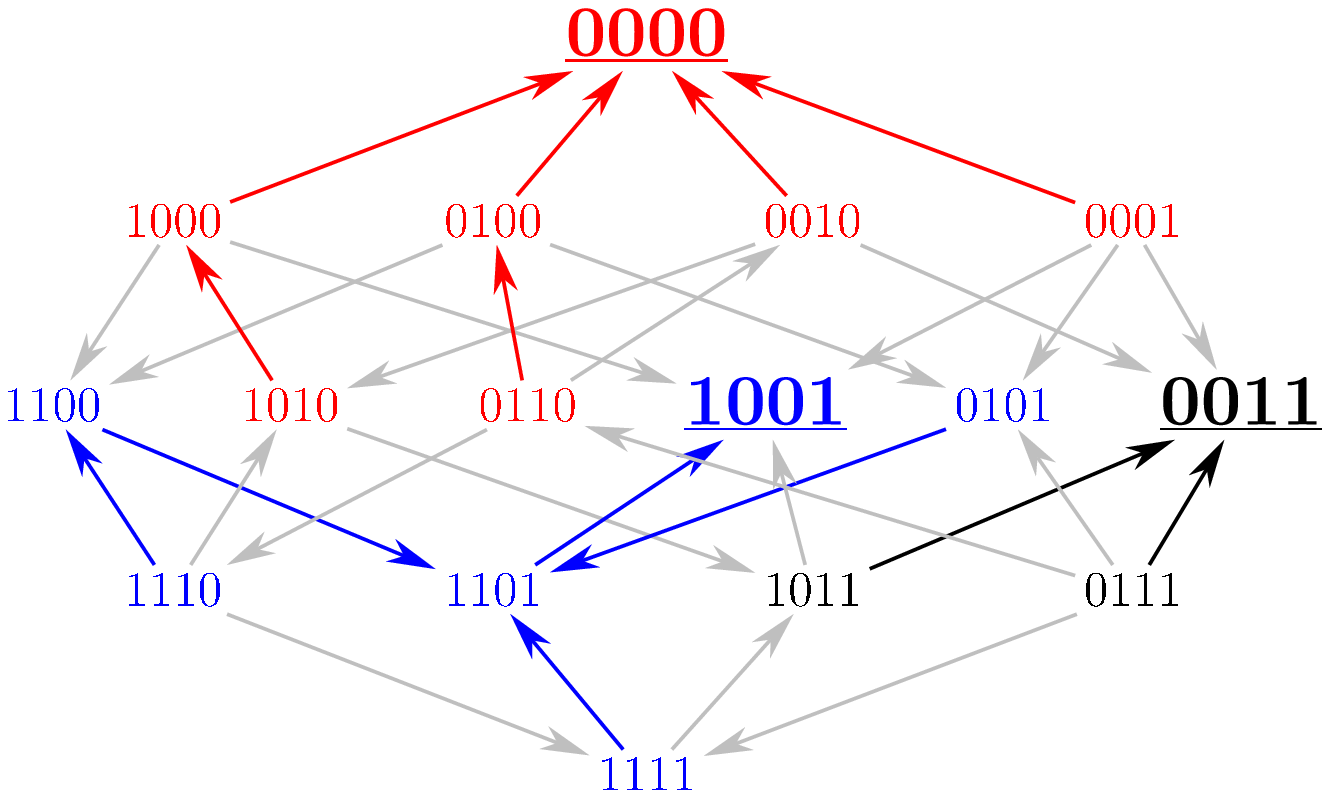}
     \caption{Fitness graphs for $L=3$ (left) and $L=4$ (right) loci. The graphs are $L$-dimensional hypercubes, and the arrows on the links point in the direction of increasing fitness. In the left panel the
     fitness values are additionally indicated by the size of the
     balls surrounding each node. The right panel shows the
     experimentally determined fitness values of all combinations of a
     subset of 4 mutations taken from the 
     8-dimensional fitness landscape displayed in
     Fig.~\ref{Fig:Aniger}. Genotypes which constitute local fitness maxima are shown in larger
     font and underlined. Colored arrows show the steps taken by a greedy dynamics,
     where the beneficial mutation of largest effect is chosen
     deterministically \cite{Franke}.}
     \label{Fig:Epistasis2}
\end{figure}

The set of binary sequences of length $L$ can be represented
graphically by linking sequences that differ at a single site. The
resulting graphs are $L$-dimensional (hyper)cubes, which are shown for
$L=3$ and $L=4$ in Fig.~\ref{Fig:Epistasis2}. Unlike the
two-dimensional case depicted in Fig.~\ref{Fig:Epistasis}, fitness
functions defined on hypercubes with $L \geq 3$ cannot be easily
plotted. A useful representation that retains partial
information about the ordering of fitness values is provided by
\textit{fitness graphs}\iffindex{fitness graph}, where the links between genotypes are
decorated with arrows pointing in the direction of increasing
fitness \cite{deVisser2009,Crona}. For more than 6 loci also this representation becomes
unwieldy and any graphical rendering of the fitness landscape is bound
to obscure at least some its geometrical structure (Fig.~\ref{Fig:Aniger}).

\begin{figure}[ht]
    \centering
\includegraphics[width=0.8\hsize]{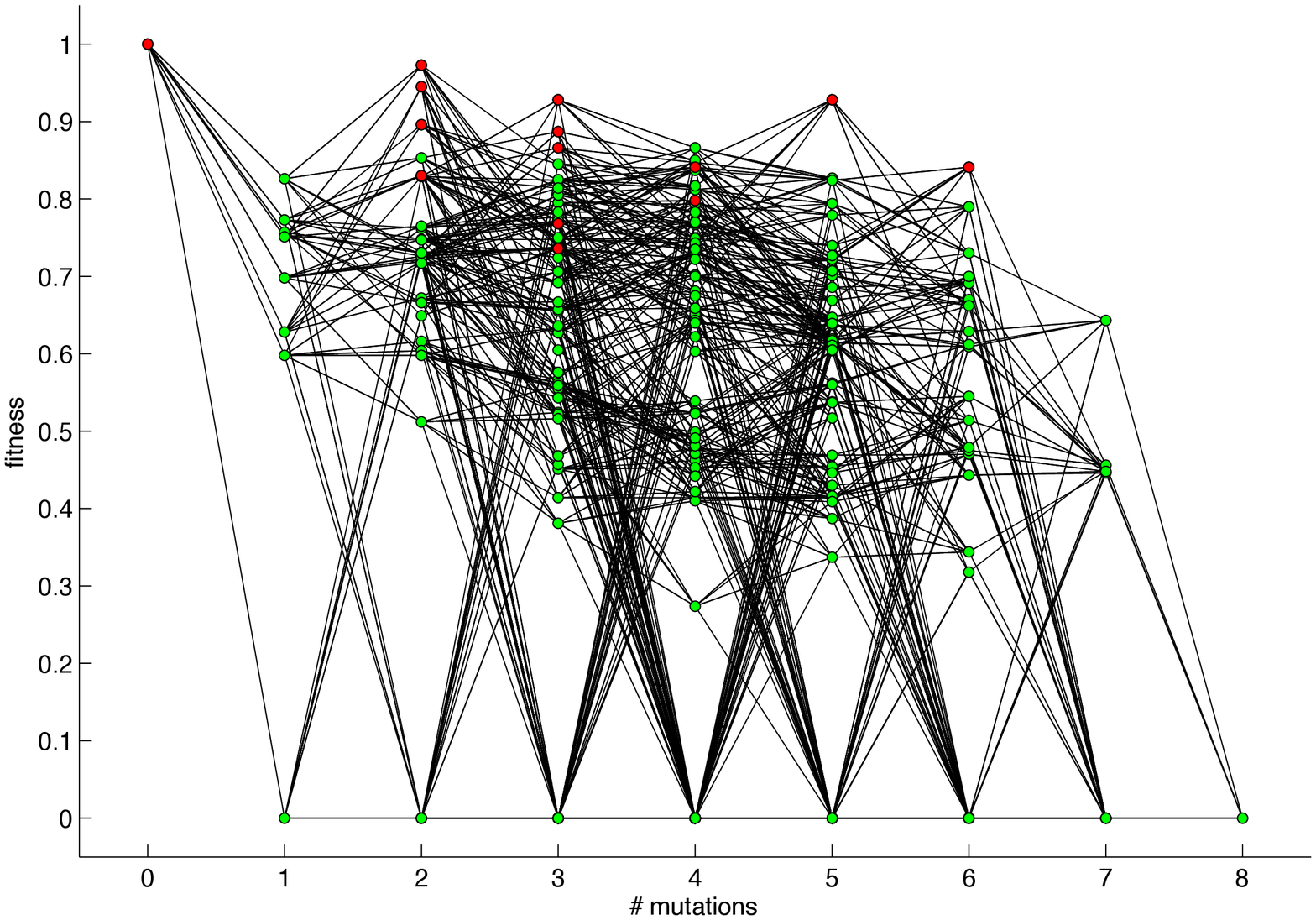}
     \caption{Fitness landscape composed of 8 individually deleterious
       marker mutations in the filamentous fungus {\emph{Aspergillus
           niger}} \cite{deVisser2009,Franke}. Each mutation resides
       on a different chromosome, and combinations of mutations were
       generated by exploiting the parasexual cycle of the
       fungus. Fitness was measured in terms of mycelial growth rate
       and normalized to the maximal (wild type) growth rate. Out of
       the $2^8=256$ possible combinations, 70 were not detected in
       the experiment. The corresponding genotypes were therefore
       classified as lethal and assigned zero fitness. The fitness
       values of the $8 \choose k$ genotypes that carry the same
       number $k$ of mutations are plotted above the same point of the
       horizontal axis, and the fitness values of genotypes that
       differ by a single mutation are connected by lines. Local
       fitness maxima are indicated in red. Courtesy of Ivan G. Szendro.}
     \label{Fig:Aniger}
\end{figure}

\subsection{Mutational pathways}

We are now ready to integrate the elements of evolutionary dynamics developed in Sects.~\ref{Sec:Moran} and \ref{Sec:Regimes} with the fitness landscape picture.  For this, we assign the $2^L$ 
genotypes of the fitness landscape to a population of $N$ individuals evolving according to the Moran model. In the periodic selection regime described in Sec.~\ref{Sec:CI} a simple
dynamics emerges. Most of the time the population is genetically homogeneous and hence occupies a single site in the fitness graph. Occasionally beneficial mutations arise 
and fix, which implies that the population moves to a neighboring
genotype of higher fitness. The population thus moves across the
fitness graph following the direction of the arrows on the links, and
we see that the fitness graph provides a kind of road map of possible 
evolutionary trajectories. 

Within the periodic selection regime the population is constrained to evolve along pathways of monotonically increasing fitness, and such paths have been termed
(evolutionarily) accessible \cite{Franke,Weinreich2006}. It is of interest to ask how likely accessible pathways are to be found on real fitness landscapes. Consider two binary genotypes 
that differ at $D$ loci. To evolve from one to the other, mutations have to take place at $D$ sites, and \textit{ a priori} these mutations can occur in any one out of $D!$ orderings. Thus the total number
of mutational pathways connecting the two genotypes is $D!$. In  a seminal experimental study, Weinreich and collaborators considered the pathways along which a highly resistant 
five-fold mutant of the TEM-1 $\beta$-lactamase enzyme could evolve
\cite{Weinreich2006} \iffindex{antibiotic resistance}
(see Fig.~\ref{Fig:Parallel} for further information on this system). They found that only 18 out of $5!=120$ pathways displayed monotonically
increasing resistance, and hence could be considered accessible under a process in which one mutation fixes at a time. Since each accessible pathway consists of a sequence of fixation
events, the weight of a pathway can be further quantified by the product of the relative fixation probabilities (\ref{qk}) along the path. Under this measure only a handful of dominating
pathways were identified for the $\beta$-lactamase system, leading the
authors to the conclusion that Darwinian evolution is much more
constrained, and hence predictable, than previously recognized.     

The periodic selection dynamics terminates at local fitness maxima, which are sinks in the fitness graph where all links carry incoming arrows (Fig.~\ref{Fig:Epistasis2}). As no beneficial 
mutations are available at such a
genotype, the population cannot escape. In practice this implies that higher order processes such as double mutations or stochastic tunneling events become important, which occur
on time scales beyond those considered in the periodic selection scenario \cite{Weinreich2005a,Szendro2013}. Local fitness maxima limit the accessibility of high fitness genotypes and present obstacles
to the evolutionary process, a concern that was articulated already by Sewall Wright in the pioneering paper which first introduced the fitness landscape concept \cite{Wright2}.  
The number of local fitness maxima and the number of evolutionarily
accessible pathways leading up to them constitute important measures of fitness landscape complexity, which have been applied
to a wide range of empirical data sets over the past few years \cite{deVisserKrug,Szendro}.

\vspace*{0.5cm}

\par{\bf Acknowledgements:} My understanding of evolutionary phenomena has been shaped by many enjoyable collaborations which were funded by DFG within SFB 680 {\em Molecular bases
of evolutionary innovations} and SPP 1590 {\em Probabilistic
structures in evolution}. Special thanks are due to Arjan de Visser
and his group at Wageningen University.  




\end{document}